\documentclass[a4paper,11pt]{article}
\usepackage{a4wide}
\usepackage{amsmath,amssymb,amsfonts,amsthm}
\usepackage[all]{xy}
\usepackage{color}

\usepackage{amscd}

\def\D{\hbox{D\kern-.73em\raise.25ex\hbox{-}\raise-.25ex\hbox{ }}}
 \def\d{\hbox{d\kern-.33em\raise.75ex\hbox{-}\raise-.75ex\hbox{}}}

\def\ed{\end{document}}
\def\beq{\begin{equation}}
\def\eeq{\end{equation}}
\def\bea{\begin{eqnarray}}
\def\eea{\end{eqnarray}}
\def\ba{\begin{array}}
\def\ea{\end{array}}
\def\bi{\begin{itemize}}
\def\ei{\end{itemize}}

\def\nn{\nonumber}

\newcommand{\bp}{\noindent\begin{minipage}[c]}
\newcommand{\ep}{\end{minipage}}

\begin{document}

\title
{Genetic Code and Number Theory}
\author{{Branko Dragovich} \\ {} \\
{Institute of Physics, Pregrevica 118, 11080 Zemun, Belgrade,
Serbia}}

\maketitle




\begin{abstract}
Living organisms are the most complex, interesting and significant
objects regarding all substructures of the universe. Life science
is regarded as a science of the 21st century and one can expect
great new discoveries in the near futures. This article contains
an introductory brief review of genetic information, its coding
and translation of genes to proteins through the genetic code.
Some theoretical approaches to the modelling of the genetic code
are presented. In particular, connection of the genetic code with
number theory is considered and the role of $p$-adic numbers is
underlined.

\end{abstract}

\maketitle

\section{Introduction}

Francis Crick (1916--2004), who together with James Watson
(1928--) discovered double helicoidal structure of DNA, in 1953
announced ``We have discovered the secret of life'' \cite{Hayes}.
However, if it was a secret of life, then life has still many
secrets.  One of them is the genetic code. Although genetic code
was finally experimentally deciphered in 1966, its theoretical
understanding has remained unsatisfactory and new models have been
offered from time to time. The genetic code is still a subject of
 more or less intensive investigation from mathematical,
physical, chemical, biological and bioinformation point of view.

It is worth recalling the emergence of special theory of
relativity and quantum mechanics. They both appeared as a result
of unsatisfactory attempts to extend classical theory to new
physical phenomena, invention of appropriate new physical concepts
and use of suitable new mathematical methods. Although far from
everyday experience these two new theories describe physical
reality quite successfully. We believe that a similar situation
should happen in theoretical description of living processes in
biological organisms. To this end, ultrametric and $p$-adic
methods seem to be very promising tools in further investigation
of life.

Here we want to emphasize the role of ultrametric distance, and in
particular, $p$-adic one. Namely, some parts of a biological
system can be considered simultaneously with respect to different
metrics -- the usual Euclidean metric, which measures spatial
distances, and some other metrics, which measure nearness related
to some bioinformation (or other) properties.

The general notion of metric space $(M,d)$ is introduced in 1906
by Maurice Fr\'echet (1878--1973), where $M$ is a set and $d$ is a
distance function. Distance $d$ is a real-valued function of any
two elements $x,y \in M$ which must satisfy the following three
properties:
\begin{align} \label{1.1}
&(i) \,\, d(x,y) = 0 \Leftrightarrow x=y,\\ &(ii) \,\, d(x,y) =
d(y,x), \\ &(iii) d(x,y)\,\, \leq d(x,z) + d(z,y),
\end{align}
where last property is called triangle inequality. An ultrametric
space is a metric space which satisfies  strong triangle
inequality, i.e.
\begin{align}
d(x,y) \leq \text{max} \{d(x,z),d(z,y) \}.
\end{align}
Word ultrametric is introduced in 1944 by Marc Krasner
(1912--1985), although examples of ultrametric spaces have been
known earlier under different names. An important class of
ultrametric spaces contains fields of $p$-adic numbers, which are
introduced in 1897 by Kurt Hensel (1861--1941). Taxonomy, which
started 1735 by Carl Linn\'e (1707--1778) as biological
classification with hierarchical structure, is another significant
example of ultrametricity \cite{ultrametricity}.

In this article we consider some aspects of the genetic code using
an ultrametric space, which elements are codons presented with
some natural numbers and the distance between them is the $p$-adic
one. However, to have a self-contained and comprehensible
exposition of the genetic code and its connection with number
theory, we shall first briefly review some basic notions from
molecular biology.

\section{Some Notions of Molecular Biology}

One of the essential characteristics that differentiate a living
organism from all other material systems is related to its genome.
The genome of an organism is its entire  hereditary information
encoded in the desoxyribonucleic acid (DNA), and contains  both
genes and non-coding sequences. In some viruses genetic material
is encoded in the ribonucleic acid (RNA). Investigation of the
entire genome is the subject of genomics.

The DNA  is a macromolecule composed of two polynucleotide chains
with a double-helical structure. Nucleotides consist of a base, a
sugar and a phosphate group. Helical backbone is a result of the
sugar and phosphate groups. There are four bases and they are
building blocks of the genetic information. They are called
adenine (A), guanine (G), cytosine (C) and thymine (T). Adenine
and guanine are derived from purine, while cytosine and thymine
from pyrimidine. In the sense of information, the nucleotide and
its base represent the same object. Nucleotides are arranged along
chains of double helix through base pairs A-T and C-G bonded by 2
and 3 hydrogen bonds, respectively. As a consequence of this
pairing there is an equal number of cytosine and guanine as well
as the equal rate of adenine and thymine. DNA is packaged in
chromosomes, which are localized in the nucleus of the eukaryotic
cells.

The main role of  DNA is to store genetic information and there are
two main processes to exploit this information. The first one is
replication, in which  DNA duplicates giving two new DNA containing
the same information as the original one. This is possible owing to
the fact that each of two chains contains complementary bases of the
other one. The second process is related to the gene expression,
i.e. the passage of DNA gene information to proteins. It is
performed by the messenger ribonucleic acid (mRNA), which is usually
a single polynucleotide chain. The mRNA is synthesized during the
first part of this process, known as transcription, when nucleotides
C, A, T, G from DNA are respectively transcribed into their
complements G, U, A, C in mRNA, where T is replaced by U (U is the
uracil, which is a pyrimidine). The next step in gene expression is
translation, when the information coded by codons in the mRNA  is
translated into proteins. In this process   transfer tRNA and
ribosomal rRNA also participate.

Codons are ordered  trinucleotides composed of C, A, U (T) and G.
Each of them presents  information which controls use of one of the
20 standard amino acids or stop signal in synthesis of proteins.

Protein synthesis in all eukaryotic cells is performed in the
ribosomes of the cytoplasm. Proteins \cite{finkelshtein} are
organic macromolecules composed of amino acids arranged in a
linear chain. Amino acids \cite{wiki} are molecules that consist
of amino, carboxyl and R (side chain) groups. Depending on R group
there are 20 standard amino acids. These amino acids are joined
together by a peptide bond. Proteins are substantial ingredients
of all living organisms participating in various processes in
cells and determining the phenotype of an organism.  The study of
proteins, especially their structure and functions, is called
proteomics. The proteome is the entire set of proteins in an
organism.

The human genome, which presents all genetic information of the
{\it Homo sapiens}, is composed of about $3 \cdot 10^9$ DNA base
pairs and contains about  $3 \cdot 10^4$ genes \cite{watson}. In
the human body there may be about 2 million different proteins.
The sequence of amino acids in a protein is determined by the
sequence of codons contained in the corresponding DNA gene. After
transcription of a gene from DNA to mRNA there is a maturation of
the primary sequence of codons to the final one which determine
primary structure of the corresponding protein. Thus not only DNA
but also RNA play important role in the gene expression. For more
detailed and comprehensive information on molecular biology  and
the genetic code one can refer to \cite{watson,matic}.

\section{Genetic Code}

The relation between codons and amino acids is known as the {\it
genetic code} \cite{wiki1}. From mathematical point of view, the
genetic code is a map from the set of 64 codons to the set of 20
amino acids and one stop signal.

So far there are about 20 known versions of the genetic code (see,
e.g. \cite{osawa}), but the most important are two of them: the
standard  code and the vertebral mitochondrial code.

In the sequel we shall mainly have in mind the vertebral
mitochondrial code, because it is a simple one and the others may
be regarded as its slight modifications.  There are $4 \times
4\times 4 = 64$  codons. In the vertebral mitochondrial code, $60$
of codons are related to the $20$ different amino acids and $4$
stop codons make termination signals. According to experimental
observations, two amino acids are coded by six codons, six amino
acids by four codons, and twelve amino acids by two codons. This
property that some amino acids are coded by more than one codon is
known as {\it genetic code degeneracy}. This degeneracy is a very
important property of the genetic code and gives an efficient way
to minimize errors caused by mutations and translation.

There is in principle up to $21^{64}$ of all possible mappings
from $64$ codons to 20 amino acids and one stop signal. It is
obvious that some of them cannot ply role of the genetic code.
Since there is still a huge number of possibilities for genetic
codes and only a very small number of them is represented in
living cells, it has been a persistent theoretical challenge to
find an appropriate approach explaining about 20 contemporary
genetic codes.

The first genetic model was proposed in 1954 by physicist George
Gamow (1904--1968), which he called the diamond code. In his model
codons are composed of  three nucleotides and proteins are
directly synthesized at DNA: each cavity at DNA attracts one of 20
amino acids. This is  an overlapping code and was ruled out by
analysis of correlations between amino acids in proteins. The next
model of the genetic code was proposed in 1957 by Crick, and is
known as the comma-free code. This model was so elegant that it
was almost universally accepted. However, an experiment in 1961
demonstrated that UUU codon codes amino acid phenylalanine, while
by the comma-free code it codes nothing. Gamow's and Crick's
models are very pretty but wrong -- living world prefers actual
codes, which are more stable with respect to possible errors (for
a popular review of the early models, see \cite{Hayes}).

An intensive study of the connection between ordering of
nucleotides in  DNA (and RNA) and ordering of amino acids in
proteins led to the experimental deciphering of the standard
genetic code in the mid-1960s. The genetic code is understood as a
dictionary for translation of information from  DNA (through RNA)
to synthesis of proteins by amino acids. The information on amino
acids is contained in codons: each codon codes either an amino
acid or termination signal (see, e.g.  a table of the vertebral
mitochondrial code). To the sequence of codons in RNA corresponds
quite definite sequence of amino acids in a protein, and this
sequence of amino acids determines primary structure of the
protein.

At the time of deciphering, it was mainly believed that the
standard code is unique, result of a chance and fixed a long time
ego. Crick \cite{crick} expressed such belief in his "frozen
accident" hypothesis, which has not been supported by later
observations. Moreover, so far at least 20  different codes have
been discovered and some general regularities found. At first
glance the genetic code looks rather arbitrary, but it is not.
Namely, mutations between synonymous codons give the same amino
acid. When mutation alters an amino acid then it is like
substitution of the original by a similar one. In this respect the
code is almost optimal.

Despite of  remarkable experimental successes, there is  no simple
and generally accepted theoretical understanding of the genetic
code. There are many papers in this direction, scattered in
various journals, with theoretical approaches based more or less
on chemical, physical, biological and mathematical aspects of the
genetic code. However, the foundation of biological coding is
still an open problem. In particular, it is not clear why genetic
code exists just in  few known ways and not in many other possible
ones. What is a principle (or principles) employed in
establishment of a basic (mitochondrial) code? What are properties
of codons connecting them into definite multiplets which code the
same amino acid or termination signal?

Let us mention some models of the genetic code after deciphering
standard code. In 1966 physicist Yuri Rumer (1901--1985)
emphasized the role of the first two nucleotides in the codons
\cite{rumer}. There are models which are based on chemical
properties of amino acids (see, e.g. \cite{swanson}). In some
models   connections between  number of constituents of amino
acids and nucleotides and some properties of natural numbers are
investigated (see \cite{rakocevic,scherbak} and references
therein). A model based on the quantum algebra $\mathcal{U}_q
(sl(2)\oplus sl(2))$ in the $q \to 0$ limit was proposed as a
symmetry algebra for the genetic code (see \cite{sorba1} and
references therein). In a sense this approach mimics quark model
of baryons. Besides some successes of this approach, there is a
problem with rather many parameters. There are also papers (see,
e.g. \cite{hornos}, \cite{forger} and \cite{bashford}) starting
with 64-dimensional irreducible representation of a Lie
(super)algebra and trying to connect multiplicity of codons with
irreducible representations of subalgebras arising in a chain of
symmetry breaking. Although interesting as an attempt to describe
evolution of the genetic code  these Lie algebra approaches did
not progress further.   For a very brief review of these and some
other theoretical approaches to the genetic code one can see
\cite{sorba1}. There is still no generally accepted explanation of
the genetic code.

\bigskip

\section{Some Mathematical Preliminaries and $p$-Adic Codon
Space}

As a new tool to study the Diophantine equations, $p$-adic numbers
are introduced by German mathematician Kurt Hensel in 1897. They
are involved in many branches of modern mathematics. An elementary
introduction to $p$-adic numbers can be found in the book
\cite{gouvea}. However, for our purposes we will use here only a
small portion of $p$-adics, mainly some finite sets of integers
and ultrametric distances between them.

Let us introduce the  set of natural numbers

\beq \mathcal{C}_5\, [64] = \{ n_0 + n_1\, 5 + n_2\, 5^2 \,:\,\, n_i
= 1, 2, 3, 4 \}\,,   \label{2.1}\eeq where $n_i$ are digits related
to nucleotides by the following assignments: C (cytosine) = 1,\, A
(adenine) = 2,\, T (thymine) = U (uracil) = 3,\, G (guanine) = 4.
This is a finite expansion to the base $ 5$. It is obvious that $5$
is a prime number and that the set $\mathcal{C}_5 [64]$ contains
$64$ numbers between $31$ and $124$ in the usual base $10$. In the
sequel we shall often denote elements of $\mathcal{C}_5 [64]$ by
their digits to the base $5$ in the following way: $ n_0 + n_1\, 5 +
n_2\, 5^2 \, \equiv n_0\, n_1\, n_2$. Note that here ordering of
digits is the same as in the expansion, i.e this ordering is
opposite to the usual one. There is now an evident one-to-one
correspondence between codons in three-letter  notation and number
$n_0\, n_1\, n_2$ representation.

There is no summation, subtraction, multiplication and division on
the codon space. A mapping of codons to codons is possible by
replacement of a nucleotide by another. In other words, there is a
sense interchange of digits  on the space $\mathcal{C}_5 \, [64]$,
but not standard arithmetic operations (summation, subtraction,
multiplication and division).

It is also often important to know a distance between numbers.
Distance can be defined by a norm. On the set $\mathbb{Z}$ of
integers  there are two kinds of nontrivial norm: usual absolute
value $|\cdot|_\infty$ and $p$-adic absolute value $|\cdot|_p$ ,
where $p$ is any prime number. The usual absolute value is well
known from elementary  mathematics and the corresponding distance
between two numbers $x$ and $y$ is $d_\infty (x, y) =
|x-y|_\infty$.

The $p$-adic absolute value is related to the divisibility of
integers by prime number $p$. Difference of two integers is again
an integer. $p$-Adic distance between two integers can be
understood as a measure of  divisibility by $p$ of their
difference (the more divisible, the shorter). By definition,
$p$-adic norm of an integer $m \in \mathbb{Z}$, is $|m|_p =
p^{-k}$, where $ k \in \mathbb{N} \bigcup \{ 0\}$ is degree of
divisibility of $m$ by prime $p$ (i.e. $m = p^k\, m'\,,$  where $
m'$ is not divisible by $p$)  and $|0|_p =0.$ $\mathbb{N}$ and
$\mathbb{Z}$ are the set of natural numbers and the set of
integers, respectively. This norm is a mapping from $\mathbb{Z}$
into non-negative rational numbers and has the following
properties:

(i) $|x|_p \geq 0, \,\,\, |x|_p =0$ if and only if $x = 0$,

(ii) $|x\, y|_p = |x|_p \,  |y|_p \,,$

(iii) $|x + y|_p \leq \, \mbox{max}\, \{ |x|_p\,, |y|_p \} \leq
|x|_p + |y|_p $ for all $x \,, y \in \mathbb{Z}$.

\noindent Because of the strong triangle inequality $|x + y|_p
\leq \, \mbox{max} \{ |x|_p\,, |y|_p \}$, $p$-adic absolute value
belongs to non-Archimedean (ultrametric) norm. One can easily
conclude that $0 \leq |m|_p \leq 1$ for any $m\in \mathbb{Z}$ and
any prime $p$.

$p$-Adic distance between two integers $x$ and $y$ is
\begin{equation}
d_p (x\,, y) = |x - y|_p \,.    \label{2.2}
\end{equation}
Since $p$-adic absolute value is ultrametric, the $p$-adic distance
(\ref{2.2}) is also ultrametric, i.e. it satisfies
\begin{equation}
d_p (x\,, y) \leq\, \mbox{max}\, \{ d_p (x\,, z) \,, d_p (z\,, y) \}
\leq d_p (x\,, z) + d_p (z\,, y) \,, \label{2.3}
\end{equation}
where $x, \, y$ and $z$ are any three integers.

The above introduced set $\mathcal{C}_5\, [64]$ endowed by
$p$-adic distance we shall call {\it $p$-adic codon space}, i.e.
elements of $\mathcal{C}_5\, [64]$ are codons denoted by $n_0 n_1
n_2$. $5$-Adic distance between two codons $a, b \in \mathcal{C}_5
\, [64]$ is

\beq d_5 (a,\, b) = |a_0 + a_1 \, 5 + a_2 \, 5^2 - b_0 - b_1 \, 5
- b_2 \, 5^2 |_5 \,,   \label{2.4} \eeq where $a_i ,\, b_i \in \{
1, 2, 3, 4\}$. When $a \neq b$ then $d_5 (a,\, b)$ may have three
different values:
\begin{itemize}
\item $d_5 (a,\, b) = 1$ if $a_0 \neq b_0$, \item $d_5 (a,\, b) =
1/5$ if $a_0 = b_0 $ and $a_1 \neq b_1$, \item $d_5 (a,\, b) =
1/5^2$ if $a_0 = b_0 \,, \,\,a_1 = b_1$ and $a_2 \neq b_2 $.
\end{itemize}
 We
see that the largest $5$-adic distance between codons is $1$ and
it is the maximum $p$-adic distance on $\mathbb{Z}$. The smallest
$5$-adic distance on the codon space is $5^{-2}$.

If we apply real (standard) distance $d_\infty (a,\, b) = |a_0 +
a_1 \, 5 + a_2 \, 5^2 - b_0 - b_1 \, 5 - b_2 \, 5^2 |_\infty $,
then third nucleotides $a_2$ and $b_2$ would play more important
role than those at the second position (i.e $a_1 \, \mbox{and} \,
b_1$), and nucleotides $a_0$ and $b_0$ are of the smallest
importance. In real $\mathcal{C}_5 [64]$ space distances are also
discrete, but take values $1,\, 2,\, \cdots , 93$. The smallest
real and the largest $5$-adic distance are equal to $1$. While
real distance describes spatial separation, this $p$-adic one
serves to describe information nearness  on the codon space.

It is worth emphasizing that the metric role of digits depends on
their position in number expansion and it is quite opposite  in
real and $p$-adic cases.  We shall see later that the first two
nucleotides in a codon are more important than the third one and
that $p$-adic distance between codons is a natural one in
description of their information content (the nearer, the more
similar meaning).

\bigskip

\section{$p$-Adic Genetic Code}

Modelling of the genetic code, the genome and proteins  is a
challenge as well as an opportunity for application of $p$-adic
distances. Recently \cite{dragovich1,dragovich1a,dragovich1b}, it
was introduced and considered a $p$-adic approach to DNA and RNA
sequences, genome and  the genetic code. The central point of this
approach is an appropriate identification of four nucleotides with
digits $1,\, 2, \, 3, \, 4$ of $5$-adic representation of some
positive integers and application of $p$-adic distances between
obtained numbers. $5$-Adic numbers with three digits form $64$
integers which correspond to  $64$ codons. It is unappropriate to
use the digit $0$ for a nucleotide because it leads to
non-uniqueness  in representation of the codons by natural
numbers. For example, $123 = 123000$ as numbers, but $123$ would
represent one and $123000$ two codons. This is also a reason why
we do not use $4$-adic representation for codons, since it would
contain a nucleotide presented by digit $0$.  One can use $0$ as a
digit to denote absence of any nucleotide. As one of the main
results that we have obtained is explanation of the structure of
the genetic code degeneracy using $p$-adic distance between
codons. A similar approach to the genetic code was later
considered on diadic plane \cite{kozyrev}, and recently
\cite{kozirev1} $2$-adic distance was applied to the PAM matrix in
bioinformatics.

 Let us mention that $p$-adic models
in mathematical physics have been actively considered since 1987
(see \cite{freund}, \cite{vladimirov1} for early reviews and
\cite{dragovich2,dragovich3,dragovich4} for some recent reviews).
It is worth noting that $p$-adic models with pseudodifferential
operators have been successfully applied to interbasin kinetics of
proteins \cite{avetisov3}.  Some $p$-adic aspects of cognitive,
psychological and social phenomena have been also considered
\cite{khrennikov1}.


\begin{table}
TABLE I. {Table of the vertebral mitochondrial code in the
$5$-adic and three-letter notation.\label{Tab:01}} \vspace{0.4cm} \\
\centerline{ {\begin{tabular}{|l|l|l|l|}
 \hline \ & \ & \ & \\
  111 \, CCC \, Pro &   211 \, ACC \, Thr  &  311 \, UCC \, Ser &  411 \, GCC \, Ala  \\
  112 \, CCA \, Pro &   212 \, ACA \, Thr  &  312 \, UCA \, Ser &  412 \, GCA \, Ala  \\
  113 \, CCU \, Pro &   213 \, ACU \, Thr  &  313 \, UCU \, Ser &  413 \, GCU \, Ala  \\
  114 \, CCG \, Pro &   214 \, ACG \, Thr  &  314 \, UCG \, Ser &  414 \, GCG \, Ala  \\
 \hline \  & \  &  \ & \ \\
  121 \, CAC \, His &   221 \, AAC \, Asn  &  321 \, UAC \, Tyr &  421 \, GAC \, Asp  \\
  122 \, CAA \, Gln &   222 \, AAA \, Lys  &  322 \, UAA \, Ter &  422 \, GAA \, Glu  \\
  123 \, CAU \, His &   223 \, AAU \, Asn  &  323 \, UAU \, Tyr &  423 \, GAU \, Asp  \\
  124 \, CAG \, Gln &   224 \, AAG \, Lys  &  324 \, UAG \, Ter &  424 \, GAG \, Glu  \\
 \hline \  & \  & \  &   \\
  131 \, CUC \, Leu &   231 \, AUC \, Ile  &  331 \, UUC \, Phe &  431 \, GUC \, Val  \\
  132 \, CUA \, Leu &   232 \, AUA \, Met  &  332 \, UUA \, Leu &  432 \, GUA \, Val  \\
  133 \, CUU \, Leu &   233 \, AUU \, Ile  &  333 \, UUU \, Phe &  433 \, GUU \, Val  \\
  134 \, CUG \, Leu &   234 \, AUG \, Met  &  334 \, UUG \, Leu &  434 \, GUG \, Val  \\
 \hline \ & \   & \  &   \\
  141 \, CGC \, Arg &   241 \, AGC \, Ser  &  341 \, UGC \, Cys &  441 \, GGC \, Gly  \\
  142 \, CGA \, Arg &   242 \, AGA \, Ter  &  342 \, UGA \, Trp &  442 \, GGA \, Gly  \\
  143 \, CGU \, Arg &   243 \, AGU \, Ser  &  343 \, UGU \, Cys &  443 \, GGU \, Gly  \\
  144 \, CGG \, Arg &   244 \, AGG \, Ter  &  344 \, UGG \, Trp &  444 \, GGG \, Gly  \\
\hline
\end{tabular}}{}}
\end{table}


Let us now turn to   Table I. We observe that this table can be
regarded as a big rectangle divided into 16 equal smaller
rectangles: 8 of them are quadruplets which one-to-one correspond
to 8 amino acids, and another 8 rectangles are divided into 16
doublets coding 14 amino acids and termination (stop) signal (by
two doublets at different places). There is a
 manifest symmetry in distribution of these quadruplets and
doublets. Namely, quadruplets and doublets form separately two
figures, which are symmetric with respect to the mid vertical line
(a left-right symmetry), i.e. they are invariant under interchange
$C \leftrightarrow G$  ($1 \leftrightarrow 4$) and $A
\leftrightarrow U$ ($2 \leftrightarrow 3$) at the first position
in codons at all horizontal lines. In other words, at each
horizontal line one can perform {\it doublet} $\leftrightarrow$
{\it doublet} and {\it quadruplet} $\leftrightarrow$ {\it
quadruplet} interchange around vertical midline. Recall that also
DNA is symmetric under simultaneous interchange of complementary
nucleotides $C \leftrightarrow G$ and $A \leftrightarrow T$
between its strands. All doublets in this table form a nice figure
which looks like letter $\mathbb{T}$.

It is worth noting that  the above invariance leaves also
unchanged polarity and hydrophobicity  of the corresponding amino
acids in all but three cases: Asn $ \leftrightarrow $ Tyr, Arg $
\leftrightarrow $ Gly, and Ser $ \leftrightarrow $ Cys.

\bigskip

\subsection{Degeneracy of the genetic code}

Let us now explore distances between codons and their role in
formation of the genetic code degeneration.

To this end let us again  turn to  Table I as a representation of
the $\mathcal{C}_5\, [64]$ codon space. Namely, we observe that
there are 16 quadruplets such that each of them has the same first
two digits. Hence $5$-adic distance between any two different
codons within a quadruplet is

\bea d_5 (a,\, b) = |a_0 + a_1 \, 5 + a_2 \, 5^2 - a_0 - a_1 \, 5
- b_2 \, 5^2 |_5 \nn
\\= |(a_2 - b_2) \, 5^2|_5 = |(a_2 - b_2)|_5 \,\, | 5^2 |_5 =
5^{-2}\,, \label{2.11} \eea because $a_0 = b_0$, $a_1 = b_1$ and
$|a_2 - b_2|_5 = 1$. According to (\ref{2.11}) codons within every
quadruplet are at the smallest distance, i.e. they are nearest
compared to all other codons.

Since codons are composed of three arranged  nucleotides, each of
which is either a purine or a pyrimidine, it is natural to try to
quantify nearness inside purines and pyrimidines, as well as
distance between elements from these two groups of nucleotides.
Fortunately there is a tool, which is again related to the
$p$-adics, and now it is $2$-adic distance. One can easily see
that  $2$-adic distance between pyrimidines  C and U is $d_2 (1,
3) = |3 - 1|_2 = 1/2$ as the distance between purines  A and G,
namely $d_2 (2, 4) = |4 - 2|_2 = 1/2$. However $2$-adic distance
between C and A or G as well as distance between U and A or G is
$1$ (i.e. maximum).

  With respect to  $2$-adic distance, the above quadruplets may be regarded
 as composed of two doublets: $a = a_0\, a_1\, 1$ and $b = a_0\, a_1\, 3$
 make the first doublet, and
 $c = a_0\, a_1\, 2$ and $d = a_0\, a_1\, 4$ form the second one. $2$-Adic
 distance between codons within each of these doublets is
 $\frac{1}{2}$, i.e.
 \begin{equation}
d_2 (a,\, b) = |(3 -1)\, 5^2|_2 =\frac{1}{2} , \, \, \, d_2 (c,\,
d) = |(4 -2)\, 5^2|_2 =\frac{1}{2} ,   \label{2.12}
 \end{equation}
because $3-1 = 4 - 2 = 2$.

One can now look at Table I as a system of 32 doublets. Thus 64
codons are clustered by a very regular way into 32 doublets. Each
of 21 subjects (20 amino acids and 1 termination signal) is coded
by one, two or three doublets. In fact, there are two, six and
twelve amino acids coded by three, two and one doublet,
respectively. Residual two doublets code termination signal.

Note that 2 of 16 doublets code 2 amino acids (Ser and Leu) which
are already coded by 2 quadruplets, thus amino acids Serine and
Leucine are coded by 6 codons (3 doublets).

To have a more complete picture on the genetic code it is useful
to consider possible distances between codons of different
quadruplets as well as between different doublets. Also, we
introduce distance between quadruplets or between doublets,
especially when distances between their codons have the same
value. Thus $5$-adic distance between any two quadruplets  in the
same column is $1/5$, while such distance between  other
quadruplets is $1$. $5$-Adic distance between doublets coincides
with $5$-adic distance between quadruplets, and this distance is
$\frac{1}{5^2}$ when doublets are within the same quadruplet.

The $2$-adic distances between codons, doublets and  quadruplets
are more complex. There are three basic cases: \begin{itemize}
\item codons differ only in one digit, \item codons differ in two
digits, \item codons differ in all three digits.
\end{itemize}
In the first case, $2$-adic distance can be
$\frac{1}{2}$  or $1$ depending whether difference between digits
is $2$ or not, respectively.

Let us now look at $2$-adic distances between doublets coding
leucine and also between doublets coding serine. These are two
cases of amino acids coded by three doublets. One has the
following distances:
\begin{itemize}
\item $d_2 (332, 132) = d_2 (334, 134) = \frac{1}{2}$  for
leucine, \item $d_2 (311, 241) = d_2 (313, 243) = \frac{1}{2}$ for
serine.
\end{itemize}

If we use usual distance  between codons, instead of $p$-adic one,
then we would observe that two synonymous codons are very far, and
that those which are close code different amino acids. Thus we
conclude that not usual metric but ultrametric is inherent to
codons.

How is degeneracy of the genetic code  related to $p$-adic
distances between codons? The answer is in the following {\bf
$p$-adic degeneracy principle}: {\it Two codons have the same
meaning with respect to amino acids if they are at smallest
$5$-adic and $1/2 \,$ $2$-adic distance}. Here $p$-adic distance
plays a role of similarity: the closer, the more similar. Taking
into account all known codes (see the next subsection) there is a
slight  violation of this principle. Now it is worth noting that
in modern particle physics just broken  fundamental gauge symmetry
gives its standard model. There is a sense to introduce a new
principle (let us call it {\bf reality principle}): {\it Reality
is realization of some broken fundamental principles}. It seems
that this principle is valid not only in physics but also in all
sciences. In this context modern genetic code is an evolutionary
broken the above $p$-adic degeneracy principle.


\subsection{Evolution of the genetic code}

The origin  and early evolution of the genetic code are among the
most interesting and important  investigations  related to the
origin and  evolution of the life. However, since there are no
fossils of  organisms from that very early period of life, it
gives rise to many speculations. Nevertheless, one can  hope that
some of the hypotheses may be tested looking for their traces in
the contemporary genomes.

It seems natural to consider biological evolution as an adaptive
development of simpler living systems to more complex ones.
Namely, living organisms are open systems in permanent interaction
with environment. Thus the evolution can be modelled by a system
with given initial conditions and guided by some internal rules
taking into account environmental factors.

We are going now to conjecture on the evolution of the genetic
code using our p-adic approach to the genomic space, and assuming
that preceding  codes used simpler codons and  older amino acids.

Recall that $p$-adic codon space $\mathcal{C}_p \, \big[ (p-1)^m
\big]$ has two parameters: $p$ -- related to $p-1$ building
blocks, and $m$ -- multiplicity of the building blocks
(nucleotides) in space elements (codons).

\begin{itemize}
\item Case $\mathcal{C}_2 \, \big[ 1  \big]$ is a trivial one and
useless for a primitive code. \item Case $\mathcal{C}_3 \, \big[
2^m \big]$ with $m =1, 2, 3$ does not seem to be realistic.

\item Case $\mathcal{C}_5 \, \big[ 4^m  \big]$ with $m = 1, 2, 3$
offers a possible pattern to consider evolution of the genetic
code. Namely, the codon space could evolve in the following way:
$\mathcal{C}_5 \, \big[ 4  \big] \to \mathcal{C}_5 \, \big[ 4^2
\big] \to \mathcal{C}_5 \, \big[ 4^3  \big] = \mathcal{C}_5\, [64]
$.
\end{itemize}

\begin{table}
TABLE II {Temporal appearance of the 20 standard amino acids
\cite{trifonov}.} \vspace{0.4cm} \\  {\begin{tabular}{|llll|}
\hline \ & \ & \ & \\
(1) Glycine,  G & (2) Alanine,  A & (3) Aspartate,  D & (4) Valine,  V \\
\hline \ & \ & \ & \\
(5) Proline,  P & (6) Serine,  S & (7) Glutamate,  E & (8) Leucine,  L \\
\hline \ & \ & \ & \\
(9) Threonine,  T & (10) Arginine,  R & (11) Isoleucine,  I & (12) Glutamine,  Q \\
\hline \ & \ & \ & \\
(13) Asparagine,  N & (14) Histidine,  H & (15) Lysine,  H & (16) Cysteine,  C \\
\hline \ & \ & \ & \\
(17) Phenylalanine,  F & (18) Tyrosine,  Y & (19) Methionine,  M & (20) Tryptophan,  W \\ 
\hline
\end{tabular}}{}
\end{table}

The primary code, containing codons in the single nucleotide form
(C, A, U, G), encoded temporally appeared  the first four amino
acids \cite{trifonov}: Gly, Ala, Asp and Val (see Table II). From
the last column of Table I we conclude that the connection between
digits and amino acids is: 1 = Ala, 2 = Asp, 3 = Val, 4 = Gly. In
the primary code these digits occupied the first position in the
$5$-adic expansion, and at the next step, i.e. $\mathcal{C}_5 \,
\big[ 4 \big] \to \mathcal{C}_5 \, \big[ 4^2 \big]$, they moved to
the second position adding digits $1, 2, 3, 4$ in front of  each
of them.

It is worth noting that traces of some early peptides composed of
the first four amino acids $G, A, D,$ and $V$ have been found
recently \cite{peptides} in the form of three motifs containing
$DGD$ submotif in some present-day proteins. This is in agreement
with our conjecture on existence of the single nucleotide primary
code at the very beginning of life.

In $\mathcal{C}_5 \, \big[ 4^2  \big]$ one has 16 dinucleotide
codons which can code up to 16  amino acids. Addition of the digit
$4$ in front of already existing codons $1, 2, 3, 4$ leaves their
meaning unchanged, i.e. 41 = Ala, 42 = Asp, 43 = Val,  and 44 =
Gly. Adding digits $3, 2, 1$ in front of the primary $1, 2, 3, 4$
codons one obtains 12  possibilities for coding some new amino
acids. To decide which amino acid was encoded by which of 12
dinucleotide codons, we use as a criterion  their immutability  in
the trinucleotide coding on the $\mathcal{C}_5 \, \big[ 4^3 \big]$
space. This criterion  assumes that amino acids encoded earlier
have more stable place in the genetic code table than those
encoded later. According to this criterion we decide in favor of
the first row in each rectangle of
 Table I and result is presented in  Table III.

Transition from  dinucleotide to trinucleotide codons occurred by
attaching nucleotides $1, 2, 3, 4$ at the third position, i.e.
behind each dinucleotide. By this way one obtains new codon space
$\mathcal{C}_5 \, \big[ 4^3  \big] = \mathcal{C}_5\, [64]$, which
is significantly enlarged and  provides a pattern to generate
known contemporary genetic codes. This codon space $
\mathcal{C}_5\, [64]$ gives possibility to realize at least three
general properties of the modern code: \begin{enumerate}\item[(i)]
encoding of more than 16 amino acids, \item[(ii)] diversity of
codes, \item[(iii)] stability of the gene expression.
\end{enumerate}

Let us give some relevant clarifications.

(i) For functioning of contemporary living organisms it is
necessary to code at least 20 standard (Table II)  and 2
non-standard amino acids (selenocysteine and pyrrolysine).
Probably these 22 amino acids are also sufficient building units
for biosynthesis of all necessary contemporary proteins. While $
\mathcal{C}_5 \, \big[ 4^2 \big]$ is insufficient, the genomic
space $ \mathcal{C}_5 \, \big[ 4^3 \big]$ offers approximately
three codons per one amino acid.

(ii) The standard (often called universal) code was established
around 1966 and was thought to be universal, i.e., common to all
organisms. When the vertebral mitochondrial code was discovered in
1979, it gave rise to belief that the code is not frozen and that
there are also some other codes which are mutually different.
According to later evidence, one can say that there are at least
20 slightly different mitochondrial and nuclear codes (for a
review, see \cite{wiki1,osawa,knight} and references therein).
Different codes have some codons with  different meaning. So, in
the standard code there are the following changes in Table I:
\begin{itemize}\item 232 (AUA): Met $\rightarrow$ Ile, \item 242
(AGA) and 244 (AGG): Ter $\rightarrow$ Arg, \item 342 (UGA): Trp
$\rightarrow$ Ter.
\end{itemize}
Modifications in  20 known codes are not homogeneously distributed
on 16 rectangles of Table I. For instance, in all 20 codes codons
$4 1 i \,\,\, (i = 1, 2, 3, 4)$ have the same meaning.

(iii) Each of the 20 codes is degenerate and degeneration provides
their stability against possible mutations. In other words,
degeneration helps to minimize codon errors.

Genetic codes based on single nucleotide and dinucleotide codons
were mainly directed to code amino acids with rather different
properties. This may be the reason why amino acids Glu and Gln are
not coded in dinucleotide code (Table II), since they are similar
to Asp and Asn, respectively.   However, to become almost optimal,
trinucleotide codes have taken into account structural and
functional similarities of amino acids.

We presented here a hypothesis on the genetic code evolution taking
into account possible codon evolution, from 1-nucleotide to
3-nucleotide, and amino acids temporal appearance. This scenario may
be extended to  cell evolution, which probably should be considered
as a coevolution of all its main ingredients (for an early idea of
the coevolution, see \cite{wong}).

\bigskip

\begin{table}
TABLE III {The dinucleotide genetic code based on the $p$-adic
genomic space $\mathcal{C}_5 \, [ 4^2 ]$. Note that it encodes 15
amino acids without stop codon, but encoding serine twice.}  \vspace{0.4cm} \\
\centerline{ {\begin{tabular}{|l|l|l|l|}
 \hline \ & \ & \ & \\
 11 \, CC \, Pro &  21 \, AC \, Thr  & 31 \, UC \, Ser & 41 \, GC \, Ala  \\
  \hline \  & \  &  \ & \ \\
 12 \, CA \, His &  22 \, AA \, Asn  & 32 \, UA \, Tyr & 42 \, GA \, Asp  \\
  \hline \  & \  & \  &   \\
 13 \, CU \, Leu &  23 \, AU \, Ile  & 33 \, UU \, Phe & 43 \, GU \, Val \\
  \hline \ & \   & \  &   \\
 14 \, CG \, Arg &  24 \, AG \, Ser  & 34 \, UG \, Cys & 44 \, GG \, Gly  \\
 \hline
\end{tabular}}{}}
\end{table}

\bigskip

\section{Concluding Remarks}

There are two important aspects of the genetic code which are
related to:
\begin{enumerate}
\item[(i)]
 multiplicity of codons which code the same amino
acid,  \item[(ii)]  assignment of codon multiplets to specific
amino acids.
\end{enumerate}

The above presented $p$-adic approach gives quite satisfactory
description of the aspect (i). Ultrametric behavior of $p$-adic
distances between elements of  the $\mathcal{C}_5 \,[64]$ codon
space radically differs from the usual ones. Quadruplets and
doublets of codons  have a natural explanation within $5$-adic and
$2$-adic nearness. Degeneracy of the genetic code in the form of
doublets, quadruplets and sextuplets is a direct consequence of
$p$-adic ultrametricity between codons. $p$-Adic $\mathcal{C}_5\,
[64]$ codon space is our theoretical pattern to consider all
variants of the genetic code: some codes are direct representation
of $\mathcal{C}_5 \, [64]$ and the others are its slight evolutional
modifications.

(ii) Which amino acid corresponds to which multiplet of codons? An
answer to this question should be expected from connections
between physicochemical properties of  amino acids and anticodons.
Namely, enzyme aminoacyl-tRNA synthetase links specific tRNA
anticodon and related amino acid. Thus there is no direct
interaction between amino acids and codons, as it was believed in
Gamow's time.

Note that there are in general $4!$ ways to assign digits $1, 2,
3, 4$ to nucleotides C, A, U, G. After an analysis of all 24
possibilities, we have taken C = 1, \, A = 2,\, U = T = 3,\, G = 4
as a quite appropriate choice. In addition to various properties
already presented in this paper, the DNA base pairs  exhibit
 relation C + G = A + T = 5.

One can  express many of the above considerations on $p$-adic
information theory in linguistic terms and investigate possible
linguistic applications.

In this paper we have employed $p$-adic distances to measure
nearness between codons, which have been used to describe
degeneracy of the genetic code. It is worth noting that in other
contexts $p$-adic distances can be interpreted in quite different
meanings. For example, $3$-adic distance between cytosine and
guanine is $d_3 (1, 4) = \frac{1}{3}$, and between adenine and
thymine $d_3 (2, 3) =1$. This $3$-adic distance seems to be
natural to relate to hydrogen bonds between complements in DNA
double helix: the smaller the distance, the stronger the hydrogen
bond. Recall that C-G and A-T are bonded by 3 and 2 hydrogen
bonds, respectively.

The translation of codon sequences into proteins is highly
information-processing phenomenon. $p$-Adic information modelling
presented in this paper offers a new approach to systematic
investigation of ultrametric aspects of  DNA and RNA sequences, the
genetic code and the world of proteins. It can be embedded in
computer programs to explore the $p$-adic side of the genome and
related subjects.

The above considerations and obtained results may be regarded as
contributions  towards foundations of (i) $p$-adic theory of
information and (ii) $p$-adic theory of the genetic code.

(i) Contributions to $p$-adic theory of information contain:
\begin{itemize}
\item formulation of $p$-adic genomic space (whose examples are
spaces of nucleotides, dinucleotides and trinucleotides), \item
relation between building blocks of information spaces and some
prime numbers;
\end{itemize}

(ii) Contributions to $p$-adic theory of the genetic code include:
\begin{itemize}
\item description of codon quadruplets and doublets by $5$-adic
and $2$-adic distances, \item observation of a symmetry between
quadruplets as well as between doublets at our table of codons,
\item formulation of degeneracy principle, \item formulation of
hypothesis on codon evolution.
\end{itemize}

Many problems remain to be explored in the future on the above
$p$-adic approach to genomics. Among the most attractive and
important themes  are: \begin{itemize}\item elaboration of the
$p$-adic theory of information towards genomics and proteomics,
\item evolution of the genome and the genetic code, \item
structure and function of non-coding DNA, \item creation of the
corresponding computer programs.
\end{itemize}

\section*{Acknowledgements}
The work  on this paper was partially supported by the Ministry of
Science and Technological Development, Serbia. Author would like
to thank many colleagues for fruitful discussion, especially M.
Rako\v cevi\'c for discussions on chemical aspects of the genetic
code.

\end{document}